\documentclass[twoside,onecolumn]{article}

\usepackage[sc]{mathpazo} 
\usepackage[utf8]{inputenc}
\usepackage[T1]{fontenc}
\usepackage[english]{babel}
\usepackage[sc, small]{titlesec}    
\usepackage{nicefrac}   
\usepackage{enumitem}   

\usepackage{authblk}    

\usepackage[babel,italian=guillemets]{csquotes}
\usepackage[style=apa]{biblatex}

\usepackage{siunitx}    
\usepackage{amsmath,amssymb}    

\usepackage{graphicx} 
\usepackage{subfig} 

\newcommand{\tensor}[1]{\mathbf{#1}}

\newcommand{\trace}[1]{\text{tr}\left(#1\right)}
\newcommand{\placeholder}{\raisebox{0.4ex}{\scalebox{0.6}{$\bullet$}}}
\newcommand{\definition}{$\,:\raisebox{-0ex}{\scalebox{1}[1]{=}}\,$}

\usepackage[left=1in, right=1in, bottom=1in, top=1in]{geometry}

\title{Towards new scaling laws for wrinkling in biologically relevant fiber-reinforced bilayers}
\date{}

\author[a]{A. Mirandola}
\author[b,f]{A. Cutolo}
\author[b,f]{A. R. Carotenuto}
\author[c]{N. Nguyen}
\author[c]{L. Pocivavsek}
\author[b,f]{M. Fraldi}
\author[a,d,e,*]{L. Deseri}

\affil[a]{Department of Civil, Environmental and Mechanical Engineering, University of Trento, via Mesiano 77, 38123 Trento, Italy}
\affil[b]{Department of Structures for Engineering and Architecture, University of Napoli "Federico II", Via Claudio 21 (buildings 6-7), 80125 Napoli, Italy}
\affil[c]{Department of Surgery, Section of Vascular Surgery and Endovascular Therapy, University of Chicago, 5841 S. Maryland Av., Chicago IL 60637, USA}
\affil[d]{Department of Mechanical Engineering and Material Sciences-MEMS, Swanson School of Engineering-SSoE, University of Pittsburgh, 3700 O'Hara St., Pittsburgh PA 15261, USA}
\affil[e]{Department of Civil and Environmental Engineering, Department of Mechanical Engineering, Carnegie Mellon, 5000 Forbes Av., Pittsburgh PA 15213, USA}
\affil[f]{Laboratory of Integrated Mechanics and Imaging for Testing and Simulation (LIMITS), University of Napoli Federico II, Italy}
\affil[*]{Corresponding author. Email address: luca.deseri@unitn.it}

\begin{document}

\maketitle

\begin{abstract}
\noindent Wrinkling, creasing and folding are frequent phenomena encountered in biological and man-made bilayers made by thin films bonded to thicker and softer substrates often containing fibers. Paradigmatic examples of the latter are the skin, the brain,  and arterial walls, for which wiggly cross-sections are detected.
Although experimental investigations on corrugation of these and analog bilayers  would greatly benefit from scaling laws for prompt comparison of the wrinkling features, neither are they available nor have systematic approaches yielding to such laws ever been provided before.

This gap is filled in this paper, where a uniaxially compressed bilayer formed by a thin elastic film bonded on a hyperelastic fiber-reinforced substrate is considered. The force balance at the film-substrate interface is here analytically and numerically investigated for highly mismatched film-substrates. 
The onset of wrinkling is then characterized in terms of both the critical strain and its corresponding wavenumber. Inspired by the asymptotic laws available for neo-Hookean bilayers, the paper then provides a systematic way to achieve novel scaling laws for the wrinkling features for fiber-reinforced highly mismatched hyperelastic bilayers. Such novel scaling laws shed light on the key contributions defining the response of the bilayer, as it is characterized by a fiber-induced complex anisotropy. Results are compared with Finite Element Analyses and also with outcomes of both existing linear models and available ad-hoc scalings. Furthermore, the amplitude, the global maximum and minimum of ruga occurring under increasing compression spanning the wrinkling, period doubling and folding regimes are also obtained.
\end{abstract}

\section{Introduction}
    \label{sec:introduction}
    Corrugation is a very common geometrical feature in Nature. This is indeed the case for skin, blood vessel walls, the brain, etc. (see e.g. \textcite{genzerSoftMatterHard2006, hohlfeldUnfoldingSulcus2011, buddayWrinklingInstabilitiesSoft2017, hollandFoldingDrivesCortical2020} among many others).

\begin{figure}
    \centering
	\includegraphics{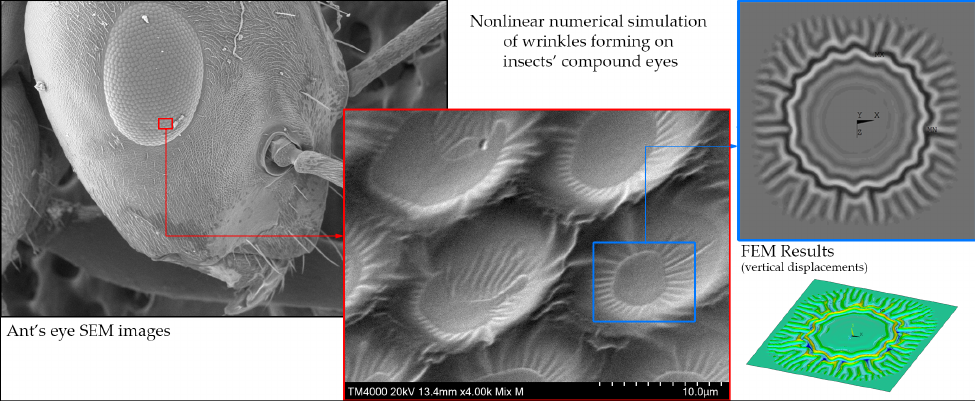}
    \caption{An original example of wrinkling in biological tissues is displayed in the images above. LEFT: A full-scale black-and-white SEM (HITACHI TM 4000 PLUS) image of the Ant's eye is reported. CENTER: A red-framed zoomed-in area from the left image is blown-up at the center of the figure: at the resolution reported in that frame, radial wrinkles are visible around each unit forming the eye's compound. \textit{Images are all original and taken by the coauthors of this paper affiliated at the LIMITS Laboratory, within the University of Napoli "Federico II"}. RIGHT: Nonlinear FE simulation. \textit{The top right} of the figure displays the projection onto the X-Z plane of the resulting displacement field representing the zoom of the blue-framed inset taken from the center of this figure: details simulated at this level of observation reveal a few azimuthal wrinkled crowns separating the central undisturbed zone from the radial wrinkles observed before at a coarser resolution. \textit{The bottom right of the figure} displays a 3D image of the vertical displacement resulting from the FE analysis, thereby reliably reproducing the experimental observation reported above.}
    \label{fig:ant-eye}
\end{figure}

\noindent For instance, as pointed out in \textcite{nguyenWrinklingInstabilitiesBiologically2020}, a wide number of papers regarding wrinkling, period doubling and quadrupling, creasing, and folding in biological systems, including tissues such as ant's eyes (see Figure~\ref{fig:ant-eye}), have been produced in the last two decades (see e.g. \textcite{genzerSoftMatterHard2006,ciarlettaPatternFormationFiberreinforced2012,benamarAnisotropicGrowthShapes2013,ciarlettaPatternSelectionGrowing2014,balbiMorphoelasticControlGastrointestinal2015, mottahediArteryBucklingAnalysis2016, gorielyMathematicsMechanicsBiological2017,alawiyeRevisitingWrinklingElastic2019,  alawiyeRevisitingWrinklingElastic2020, nathDynamicLuminalTopography2020,chenPostwrinklingBehaviorsBilayer2021, kaiMechanicalRegulationTissues2022,mostafaviyazdiMechanicalModelingCharacterization2022} and references cited therein). Furthermore, important results regarding various thin man-made mechanical systems exhibiting corrugation have been largely investigated in parallel (see e.g. \textcite{biotSurfaceInstabilityRubber1963, cerdaGeometryPhysicsWrinkling2003, pocivavsekStressFoldLocalization2008,cutoloWrinklingPredictionFormation2020} and references cited therein, among many others).

\noindent Most of the investigations mentioned above have dealt with homogeneous (hyper-) elastic bilayers, perfectly bonded to one another. Those studies have been performed primarily under either an applied prestretch or compressive in-plane external tractions. Occasionally, thermal actions or growth (with reference to biological systems), have also been analyzed as a source of possible instability through corrugation, although not so extensively. For the given action, the aforementioned literature shows that the enabling features for wrinkling are (i) the extreme thinness of one of such layers relative to the thickness of the whole system, and (ii) the mismatch of the elastic moduli of such layers. 

\noindent Unlike other phenomena, though, very few \textit{scaling laws} connecting the geometrical features of the exhibited corrugations and the mechanical properties of the bilayers described above are available. \textcolor{black} {In particular, in \textcite{allenAnalysisDesignStructural1969}} (Sect. 8.2) a slightly modified version of the scaling laws \eqref{eqn:asymptotic-law-kh-nh} and \eqref{eqn:asymptotic-law-strain-nh}, displayed in the sequel, governing the critical strain and the wavenumber at the onset of wrinkling, were obtained in a fairly simple and clever way for an elastic strut bonded to an isotropic elastic core. Such laws have been revisited in more recent times by \textcite{sunFoldingWrinklesThin2011}, where those relationships have been obtained (without showing the actual derivation) as asymptotic expansions of the analytic solutions of the wrinkling problem for isotropic hyperelastic bilayers. With regard to a completely different situation, such as free-standing thin polymeric sheets under tension, a new set of scaling laws has been provided in \textcite{cerdaGeometryPhysicsWrinkling2003}, and analytically validated (with a slight change) in \textcite{puntelWrinklingStretchedThin2011}.
More recently, in \textcite{gorielyLiquidCrystalElastomers2021}, generalizations of \eqref{eqn:asymptotic-law-kh-nh} and \eqref{eqn:asymptotic-law-strain-nh} were obtained for bilayers made of liquid crystal elastomers (with certain given initial orientations of the domains) bonded with a homogeneous and hyperelastic neo-Hookean material, both in the case in which the thin layer is neo-Hookean and the substrate is made of the liquid crystal elastomer and vice versa. 

\noindent Primarily due to the presence of fibers, \textit{scaling laws for wrinkling} occurring in biological tissues are not yet available in the literature. Nonetheless, an ad-hoc equation has been recently provided in \textcite{nguyenWrinklingInstabilitiesBiologically2020} for the critical strain at the onset of the instability, although it  did not come from  any mathematical justification. 

\noindent Among other issues, the main problem of soft biological tissues is certainly heterogeneity. This can influence the mechanical response of the tissue in terms of inhomogeneity of its pointwise elastic properties and, depending on the shape and functionality of the tissue, its residual stresses (see e.g. \textcite{TaberHumphrey2001, HaynFischerMierke2020}, and references cited therein). Nevertheless, averaging and homogenization methods yielding effective mechanical properties for biological tissues have been developed over the last two decades (for a more detailed discussion see e.g. \textcite{ROBERTSON2013275, cyron2016homogenized, braeu2017homogenized}). This leads to overall characterizations of the constitutive behavior of such complex systems
(see e.g. \textcite{HUMPHREYandRAJA2002, gasserHyperelasticModellingArterial2005, Bellini2013AMM} and ref.s cited therein) for which the degree of approximation can, of course, vary significantly  (see e.g. \textcite{ROBERTSON2013275} for a detailed discussion of this aspect). The most utilized approach for such tissues, with particular regard to arterial walls, is certainly the one introduced in \textit{\textcite{holzapfelNewConstitutiveFramework2000}} (called \textit{OGH} in the sequel). Such a constitutive equation has been utilized in \textcite{nguyenWrinklingInstabilitiesBiologically2020} and it will also be employed in the sequel together with the \textit{Standard Reinforcing Model} (called \textit{SRM} in the sequel). The latter has been studied since the eighties (see e.g. \textcite{Kurashige1981, TriantafyllidisAbeyaratne1983}), although it was later in \textcite{Qiu1997RemarksOT} that the impact of such a constitutive law on the deformation modes exhibited by this material was investigated. More recently in \textcite{melnikModelingFiberDispersion2015, SenJE2022} the SRM law has also been exploited in relation to the dispersion of the fibers.

\noindent The present paper is the first step towards finding a rigorous procedure enabling one to systematically find scaling laws for corrugation starting from the equations governing such a phenomenon. In particular, the work here is organized as follows. In Sect. \ref{sec:a-simpler-model}  the approach undertaken in \textcite{nguyenWrinklingInstabilitiesBiologically2020} for the study of wrinkling in bilayers formed by a three-dimensional stiff thin film adhering on top of an OGH (and then SRM) fiber-reinforced, and much softer and thicker, substrate is revisited through a simplified approach. %
Here, instead of treating the top layer as a three-dimensional solid, a dimensionally reduced formulation (like the plate model in \textcite{shieldBucklingElasticLayer1994}) is assumed, and the simplified constitutive SRM law for the fiber-reinforced substrate is considered. 
 
 \noindent In Sect. \ref{sec:an-asymptotic-law}, a comparison of the outcomes of choosing SRM instead of the more complex OGH law is performed.
 Indeed, such a comparison is produced for the analytic results for both the critical strain and the wavenumber at the onset of wrinkling coming from the OGH constitutive law both by considering the top layer as (i) a three-dimensional solid and (ii) as a plate, and (iii) the SRM law for such a dimensionally reduced approach. 
 
  \noindent Furthermore, asymptotic expansions for both the wrinkling strain and the corresponding wave number have been provided for high-contrast elastic mismatches between the thin layer and the substrate in the presence of the reinforcing fibers. This starts from the outcome of the analytic procedure performed to seek the (a) minimum critical strain with respect to the wavenumber among the ones solving the eigenvalue problem characterizing the balance of forces at the interface between film and substrate, and (b) the corresponding wavenumber. The latter is then processed through a suitable sequence of Taylor's expansions yielding \eqref{eqn:asymptotic-law-kh},  a novel \textit{scaling law for the wavenumber} itself formed by a product of two terms, a basal one and an amplifying factor. The former term turns out to coincide with \eqref{eqn:asymptotic-law-strain-nh}, namely the asymptotic law for the wavenumber of a purely neo-Hookean bilayer reported in \textcite{sunFoldingWrinklesThin2011, caoWrinklingPhenomenaNeoHookean2012}. 
  In the cases of either the absence of the fibers or their perfect randomness, the amplifying factor goes to one, thereby letting the novel scaling law for the wavenumber degenerate to \eqref{eqn:asymptotic-law-strain-nh}. There, the wavenumber scales like the cubic root of the elastic mismatch of the two layers forming the system in that case. 
  \noindent Full novelty is instead in the amplifying factor \eqref{aplifying_factor} due to the presence of load-bearing distributed fibers within the matrix of the substrate. That factor turns out to scale with the sixth root of a sum of terms. The latter turns out to be even in the spatial dispersion of the fibers (up to the fourth power of that parameter), and modulated by suitable powers of the  modified stiffness ratio between fibers and matrix (accounting for the volume concentration of the former), and on the square of the sin of four times the relative orientation of the fibers themselves. 
  \noindent With an analog procedure, the \textit{novel scaling law} \eqref{eqn:asymptotic-law-strain} \textit{for the critical strain at the onset of wrinkling} is also obtained.
 Not surprisingly, this retrieves \eqref{eqn:asymptotic-law-strain-nh} (see \textcite{sunFoldingWrinklesThin2011,caoWrinklingPhenomenaNeoHookean2012}) for isotropic neo-Hookean bilayers, either when fibers are absent or whenever they are randomly distributed. In all of the other cases, the modulating function arising in \eqref{eqn:asymptotic-law-strain}
 depends on the presence of the fibers and it is nothing but the square of the amplifying factor previously obtained for the wavenumber. In this same section, diagrams showing the comparisons between the obtained scalings, the analytic results obtained in the previous section, and numerical results performed by using ABAQUS for Finite Element Methods (FEM) simulations have been displayed. Such figures relate to results for high elastic contrast between the top thin layer and the substrate and given sets of parameters, carefully discussed in Sect.  \ref{sec:an-asymptotic-law}.

 \noindent Finally, in Sect. \ref{sec:orthotropic} a re-interpretation of the obtained asymptotic expansions for both the critical strain and the associated wavenumber is proposed in terms of the resulting properties of the linearized system about the underformed state  obtained in \textcite{nguyenWrinklingInstabilitiesBiologically2020}. It is worth recalling that the result of such a linearization yields an actual orthotropic material response for the substrate.
 This innovative way of looking at the newly derived scaling laws illustrates how the modulating function mentioned above is essentially related to the orthotropy of the linearized solid. Indeed, the modulating factor introduced above goes with the sixth root of a term governed by the ratio of the Young moduli evaluated in the principal system of the resulting linearized orthotropic medium, while still depending on the square of the sin of four times the relative angle between the family of the reinforcing fibers.   
    
\section{A simplified model}
	\label{sec:a-simpler-model}
	In \textcite{nguyenWrinklingInstabilitiesBiologically2020} an approach to computing the critical strain for which a thin membrane adhering to a soft substrate experiences wrinkling is presented. In that paper, the computation of such a strain is performed by considering the system as composed of two three-dimensional solids and then writing appropriate plane strain balance equations. However, this approach has the computational disadvantage of solving a highly non-linear system. In order to circumvent this drawback, the geometry and the physics of the problem suggest key simplifying assumptions leading, in a much simpler way, to almost the same results obtained from the fully three-dimensional model cited above.\\
\bigskip

A more efficient approach can be undertaken by focusing the present analysis on:
\begin{enumerate}[label=(\alph*)]
    \item\label{item:a} bilayers for which the mismatch between the elastic moduli of the layer and of the substrate is very high (i.e. between $10^4\div10^6$);
    \item\label{item:b} the layer being considered as very thin compared to the substrate (which, in mathematical terms, is in fact assumed infinitely deep). 
\end{enumerate}
\begin{figure}
    \centering
	\includegraphics{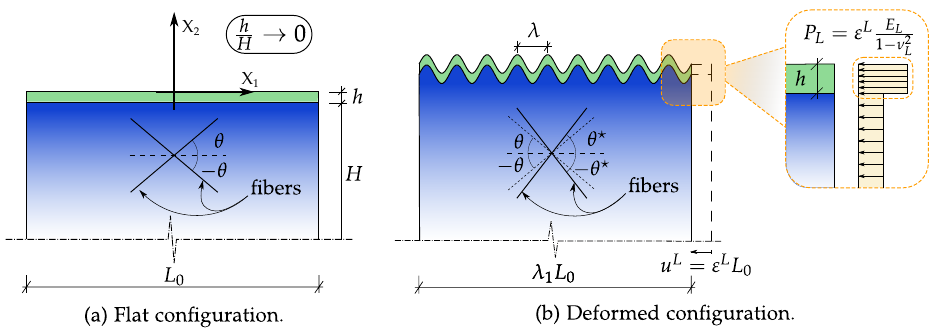}
    \caption{Schematics of the plane strain bilayer system. The substrate is composed by two families of fibers with relative angle $2\theta$ embedded in a neo-Hookean matrix. In (a) the bilayer is undeformed, with thickness $h$ and length $L_0$. This configuration is assumed to be the reference one, with the material coordinates system $X_1-X_2$, with a substrate much deeper than the layer ($h/H \to 0$). In (b) the deformed configuration is shown. There $u^L$ is an imposed contractile displacement, $\epsilon^L = u^L/L_0$ is the corresponding strain, $\lambda_1=1+\epsilon^L$ is the resulting stretch and, hence, the bilayer's deformed length is $\lambda_1 L_0$. 
    This geometry remains valid for higher values of the stretch $\lambda_1 \ge \lambda_1^{cr}$, where $\lambda_1^{cr}$ is the longitudinal stretch at the onset of wrinkling (see Eqn. \eqref{Critical Stretch}),
    spanning the whole wrinkling regime until period-doubling starts (see Fig. \ref{fig:amplitude}  in Sect. \ref{sec:an-asymptotic-law}). During deformation the angle between the two families of fibers takes the form {\small$\theta^* = 1/2 \, \cos^{-1}((\tensor{F}\tensor{l}_1 \cdot \tensor{F}\tensor{l}_2)/(|\tensor{F}\tensor{l}_1| |\tensor{F}\tensor{l}_2|))$, where $\tensor{l}_m$, $m=1,2$ are defined below \eqref{fiber directions}.} The upper left corner displays a magnification of the reactive tractions arising in the bilayer due to the imposed displacement (the tractions between the layer and the substrate are not shown to scale relative to one another).}     
    \label{fig:model}
\end{figure}
Item (b) allows for considering thin plate behavior for the top layer, and this inspired many studies on fully isotropic, homogeneous and elastic bilayers already present in the literature (see e.g. \textcite{sunFoldingWrinklesThin2011, caoWrinklingPhenomenaNeoHookean2012} and ref.s cited therein). In the present analysis, a thin plate theory to model the thin film bonded to the fiber-reinforced substrate is adopted. As previously mentioned, the latter here
is modeled through the OGH constitutive equation. Such a material has a strain energy that is additively composed of two terms. The first one is due to the classical neo-Hookean matrix. The second term is due to the presence of  fibers, organized in families, dispersed in the matrix, and reciprocally oriented with one another at a certain angle $2\theta$ (Figure~\ref{fig:model}). Finally, the total strain energy density of the substrate $W_s$ is given by the sum of those two contributions, i.e.
\begin{equation}
	\begin{gathered}
		W_{s,OGH} = W_{s,\text{matrix}} + W_{s,\text{fibers}},\\
		\text{with} \quad 
		W_{s,\text{matrix}} = \mu_M (I_{1}-3),\quad
		I_1\definition\trace{\tensor{C}},\\
		\tensor{C}\definition\tensor{F}^T\tensor{F}, \quad
		W_{s,\text{fibers}} = \dfrac{k_1}{2k_2}\sum_{m=1}^{N}\bigl[\exp\bigl(k_2 E_m^2\bigr) -1\bigr],
	\end{gathered}
	\label{eq:OGH-strain-energy}
\end{equation}
where $N$ is the number of families of fibers in the matrix and $\tensor{F}$ is the deformation gradient. The term $2\mu_M$ stands for the shear stiffness of the matrix, $k_1$ is a parameter related to the stiffness of the fibers and $k_2$ is a non-dimensional parameter determined experimentally. The term $I_1$ is the first invariant, i.e. the trace of any of the two Cauchy-Green tensors. The argument of the exponential defined above, besides $k_2$ is given by
\begin{equation}
	\begin{gathered}
		E_m = \kappa(I_{1} -3) + (1-3\kappa)(I_{4m} - 1),\\
		I_{4m} = \tensor{l}_m \cdot(\tensor{C} \tensor{l}_m) 
  \label{fiber directions}
	\end{gathered}
\end{equation}
where $\tensor{l}_m = (\cos\theta, \sin\theta, 0)^T $ is the unit vector representing the $m$-th fibers family with respect to the horizontal axis. It is worth noting that $I_{4m}$ is the magnitude (squared) of the extension/contraction of the fibers.

\noindent The outcomes of a dimensionally reduced theory for the top layer, such as the plate one adopted here, and its interactions with an OGH infinite layer have not yet been explored in the literature. Indeed, in the aforementioned recent paper by \textcite{nguyenWrinklingInstabilitiesBiologically2020}, both the layer and the OGH substrate were treated as fully three-dimensional bodies under plane-strain conditions. No matter the constitutive response of both layers nor how the film is modeled, the balance of tractions at the interface between film and substrate governs the configurations of the bilayer.

\noindent Here the configurational changes of such an interface are analyzed through a small-on-large approach, consistent with the existing literature on compressed bilayers formed by stiff films on non fiber-reinforced soft substrates (see e.g. \textcite{JiangRogersPNAS2007, caoWrinklingPhenomenaNeoHookean2012, Hutchinson2013, WangLiuYibinFu2023} and references cited therein, among many others). To this aim, following \textcite{nguyenWrinklingInstabilitiesBiologically2020}, Eqn. (4) (see e.g. also \textcite{shieldBucklingElasticLayer1994}, Eqn. (3) for the sole displacement field, and \textcite{sunFoldingWrinklesThin2011}, Eqn.s (2.1)$ \, \div $(2.3))  a sinusoidal perturbation (of amplitude $\delta\ll1$) is given to a homogeneous plane-strain, volume-preserving finite deformation of the substrate induced through a longitudinal shortening (imposed in the direction $1$, as specified below), i.e.
\begin{equation}
	\begin{gathered}
		x_1 = \lambda_1 X_1 - \delta \, \alpha \lambda_1^2 \lambda \sin(k X_1)e^{\alpha k X_2}\\
		x_2 = \lambda_2 X_2 + \delta \, \lambda \cos(k X_1) e^{\alpha k X_2}\\
		p = p_0 + \delta \, p_1 \cos(k X_1) e^{\alpha k X_2}.
	\end{gathered}
	 \label{eqn:definition-p}
\end{equation}
It is worth noting that $\lambda_2=1/\lambda_1$, $p$ is the hydrostatic pressure needed to maintain incompressibility (namely the reactive action needed to keep isochoricity of the substrate), whereas $p_0$, $p_1$ and $\alpha$ are constants (to be determined through boundary conditions). Of course, in \eqref{eqn:definition-p} the pairs $(x_1,x_2)$ and $(X_1,X_2)$ give the coordinates of a generic point in the deformed and in the reference configuration, respectively. As the wrinkling of the interface occurs, the corrugation will be characterized by a \textit{space wavelength} $\lambda$, or by its corresponding \textit{wavenumber} $k$, linked by the relation $k = 2 \pi/\lambda$.
The deformation and pressure fields \eqref{eqn:definition-p} satisfy some basic considerations about the nature of the problem. Along the direction $X_2$, the perturbation must fade at a great distance from the interface, and horizontally the motion must be periodic. Moreover, as shown further, the fields are solutions of the equilibrium of the substrate, for a suitable choice of $\alpha$ and $p_1$.\\
The incompressibility condition $\det \tensor{F} = 1$ is satisfied at the first order, i.e. $\lambda_1^{-2}\partial\,\widetilde{u_1}/\partial\,X_1 + \partial\,\widetilde{u_2}/\partial\,X_2=0$ (see e.g. \textcite{pence1991buckling,yue1994plane}). 
This last equation is identically satisfied by \eqref{eqn:definition-p} if $\widetilde{u}_i = x_i - \lambda_i X_i\; (i=1,2)$ is assumed.\\
Finally, as shown in Figure~\ref{fig:asymptotic-kh-rhoML-e-4}(e) the bifurcation is characterized by a sinusoidal profile and, by observing Figure~\ref{fig:amplitude}, this extends up to ten times the strain at the onset of the wrinkling.
\noindent Tractions acting on the film coming from the substrate must be evaluated in order to characterize which superimposed deformations are admissible for the bilayer. To do this, the first Piola-Kirchhoff stress tensor for the substrate $\tensor{P}^s$ can be computed as follows:
\begin{gather}
	\tensor{P}^{s} = \dfrac{\partial W_{s,OGH}}{\partial \tensor{F}} - p\tensor{F}^{-T}.
	\label{eq:piola-kirchhoff}
\end{gather}
When the wrinkling has not yet occurred, the constant $p_0$ can be obtained noting that the normal traction at the interface $P_{22}^s(\delta=0)$ vanishes. By letting $\delta=0$ the following expression for $p_0$ is determined, leading to
\begin{equation}
	\begin{gathered}
	p_0 =\dfrac{2\mu_M}{{\lambda_1 }^2 }+\dfrac{4k_1 {\mathrm{e}}^{k_2 {r }^2 } r {(\kappa -3 \, \kappa \, \sin^2\theta +\sin^2\theta} )}{{\lambda_1 }^2 },\\
	\text{where} \quad 
	r =\dfrac{\kappa {{({\lambda_1 }^2 -1)}}^2 }{{\lambda_1 }^2 }+\dfrac{{(3\kappa -1)}{({\lambda_1 }^2 -1)}{({\lambda_1 }^2 \, \sin^2\theta -{\lambda_1 }^2 +\sin^2\theta} )}{{\lambda_1 }^2 },
\end{gathered}
\label{eq:pressure0} 
\end{equation}
which, introduced into \eqref{eqn:definition-p} and then into \eqref{eq:piola-kirchhoff}, allows us to write the equilibrium equations for the substrate
\begin{equation}
	 P_{ij,j}^s = 0, \quad i,j = 1,2
	\label{eq:alpha}
\end{equation}
where $(\placeholder)_{,j}$ indicates $\partial(\placeholder)/\partial X_j$ and the repeated index means summation. It is worth noting that a full analytic proof of the fact that \eqref{eqn:definition-p} is a representation formula for the solution of the boundary value problem at hand for purely neo-Hookean bilayers (with no fibers) formed by stiff films on softer substrates could be provided by generalizing the approaches utilized in \textcite{pence1991buckling,Qiu1997RemarksOT,WangLiuYibinFu2023} to account for the presence of the fibers reinforcing the substrate. Of course, unlike the case of neo-Hookean materials, for fiber-reinforced substrates the constants $\alpha$ and $k$ characterizing the eigenmodes are expected to be influenced both by the fiber and by the matrix parameters.

In particular, solving equation \eqref{eq:alpha} for the systems under consideration yields four different pairs of solutions $(\alpha,p_1(\alpha))$. 
Note that the solution of $\alpha$ is formed by two complex conjugate pairs, which differ from one another with the sign of their positive part. Nevertheless, only two of those pairs $(\alpha,p_1)$ can be used, more precisely the ones that have an $\alpha$ with a strictly non-negative part, as the perturbation effects must vanish at long distances from the interface (it is worth noting that going inward deep into the substrate entails negative values for $X_2$, see fig.\eqref{fig:model}). After labeling $\alpha_1$ and $\alpha_2$ the values satisfying equations \eqref{eq:alpha}, the resulting quantities in the reference configuration are given by a linear combination of the respective eigenfunctions, i.e.
\begin{equation}
	\begin{gathered}
		\widetilde{x}_1 = C_1x_1\Big|_{\alpha=\alpha_1} + C_2x_1\Big|_{\alpha=\alpha_2} + o(\delta^2),\quad \widetilde{x}_2 = C_1x_2\Big|_{\alpha=\alpha_1} + C_2x_2\Big|_{\alpha=\alpha_2} + o(\delta^2),\\
		 \widetilde{p} = C_1p\Big|_{\alpha=\alpha_1, p_1=p_1(\alpha_1)} + C_2p\Big|_{\alpha=\alpha_2, p_1=p_1(\alpha_2)} + o(\delta^2), \\
		\widetilde{\tensor{P}}^s = C_1\tensor{P}^s\Big|_{\alpha=\alpha_1, p_1=p_1(\alpha_1)} + C_2\tensor{P}^s\Big|_{\alpha=\alpha_2, p_1=p_1(\alpha_2)} +o(\delta^2).
	\end{gathered}
	\label{eqn:eigensolutions}
\end{equation}
In the case that the stiffness of the fibers approaches zero, it is worth noting that \eqref{eq:pressure0} and \eqref{eq:alpha} give the same results found in \textcite{sunFoldingWrinklesThin2011} (note that $Q$ used in \textcite{sunFoldingWrinklesThin2011} is equal to $2\mu_M$), namely:
\begin{equation}
	\begin{gathered}
		p_0 = \dfrac{2\mu_M}{\lambda_1^2}, \quad\\
		\left(\alpha_2, p_{1}(\alpha_2)\right) = (1, 0), \\
		 \left(\alpha_1, p_{1}(\alpha_1)\right) = \left(\dfrac{1}{\lambda_1^2}, \dfrac{4\pi\mu_M(1-\lambda_1^4)}{\lambda_1^3} \right).
	\end{gathered}
	\label{eqn:eigenproblem-neo-Hookean}
\end{equation}

\noindent It is worthy of mention that the approach followed by Eqn. \eqref{eqn:definition-p} to Eqn. \eqref{eqn:eigenproblem-neo-Hookean} is analogous to the one introduced in \textcite{nguyenWrinklingInstabilitiesBiologically2020}. The assumption \ref{item:b} introduced above, i.e. the layer is assumed to be very thin compared to the substrate, is now useful. In this case it appears reasonable to assume that the layer starts to wrinkle with a wavelength that is large compared to the thickness of the upper layer. Henceforth, a plate behavior with a single bending axis which lies on a semi-infinite space can be assumed (see e.g. \parencite{shieldBucklingElasticLayer1994}). Upon utilizing the balance equation at the interface  between the top layer and the substrate (in the reference configuration) the following expressions are obtained (see \textcite{shieldBucklingElasticLayer1994, sunFoldingWrinklesThin2011})
\begin{equation}
	\begin{gathered}
	\dfrac{E_{L} h}{1-\nu_L^2} \Bigg(\dfrac{h^2}{3} \dfrac{\partial^4 \widetilde{u}_2(X_1,0)}{\partial X_1^4} - \dfrac{h}{2} \dfrac{\partial^3 \widetilde{u}_1(X_1,0)}{\partial X_1^3} + \varepsilon^L \dfrac{\partial^2 \widetilde{u}_2(X_1,0)}{\partial X_1^2}\Bigg) + \widetilde{P}_{22}^s(X_1,0) = 0\\
	\dfrac{E_{L} h}{1-\nu_L^2} \Bigg(\dfrac{\partial^2 \widetilde{u}_1(X_1,0)}{\partial X_1^2} - \dfrac{h}{2}\dfrac{\partial^3 \widetilde{u}_2(X_1,0)}{\partial X_1^3}\Bigg) - \widetilde{P}_{12}^s(X_1, 0) = 0,
    \end{gathered}
\label{eq:equilibrium-plate}
\end{equation}
where 
\begin{equation}
\varepsilon^L = 1 - \lambda_1 = \dfrac{P_L}{1-\nu_L^2} = \dfrac{u^{L}}{L_0}
\end{equation}
is the longitudinal strain in the absence of prestretch, $P_L$ is the corresponding longitudinal stress arising across the layer \parencite{shieldBucklingElasticLayer1994}, and $E_L$ and $\nu_{L}$ are the Young modulus and Poisson ratio of the layer, respectively. Furthermore, $\widetilde{P}_{ij}^s$ and $\widetilde{u}_i=\widetilde{x}_i-\lambda_iX_i,\,(i,j=1,2)$, are  the stresses exchanged between the substrate and the layer and the displacements at the interface (hence evaluated at $X_2=0$ and obtained from Eqn. \eqref{eqn:eigensolutions}), respectively. 

\noindent In order to facilitate the reader, the following notation is utilized in the sequel: \emph{M} stands for "matrix", \emph{F} for "fibers" and \emph{L} for "layer". In addition, the order reflects the position of the shear modulus of a given system within the ratio: for example, $\rho_{ML}$ means stiffness of the matrix (\emph{M}) forming the substrate over the one of the layer (\emph{L}).

\noindent Recalling that $\lambda = 2\pi / k$ denotes the spatial wavelength of periodic wrinkles, and by introducing  $k_h = 2\pi h/\lambda$, namely its corresponding \textit{non-dimensional wavenumber}, by setting $\rho_{FM}=k_1/\mu_M$ the ratio between stiffness information about both the fibers and the matrix, and by noting that $\rho_{ML}=6\mu_M/E_L$ is the stiffness ratio between the substrate and the layer, the substitution of expressions \eqref{eqn:eigensolutions} into \eqref{eq:equilibrium-plate} leads to the following homogeneous linear system in the amplitudes $C_1$ and $C_2$ appearing in \eqref{eqn:eigensolutions}:
\begin{equation}
	\tensor{M}(k_h, \lambda_1, \rho_{FM},\rho_{ML}, \kappa, \theta, k_2, \varepsilon^L)
	\begin{pmatrix}
		C_1\\C_2
	\end{pmatrix}
	= 0,
\end{equation}
where $\tensor{M}$ is the resulting coefficients matrix and the pair $C_1,C_2$ characterize the wrinkling eigenmodes. For the sake of brevity, the explicit form of $\tensor{M}$ is omitted, although available upon request. Of course, the amplitude modes $C_1$  and $C_2$ are associated with the values of $\varepsilon$ for which a bifurcation of equilibrium occurs, i.e. such that
\begin{equation}
	\varepsilon_{cr} \Big| \det\bigl(\tensor{M}(k_h, \lambda_1, \rho_{FM},\rho_{ML}, \kappa, \theta, k_2, \varepsilon^L)\bigr) = 0 \quad \Longrightarrow \quad \varepsilon_{cr} = \hat{\varepsilon}(k_{h,cr}, \lambda_1^{cr}, \rho_{FM},\rho_{ML}, \kappa, \theta, k_2),
	\label{eq:system-condition}
\end{equation}
where
\begin{equation}
    \lambda_{1}^{cr} = 1 - \varepsilon_{cr}.
    \label{Critical Stretch}
\end{equation}
It is worth noting that $\varepsilon^L$ appears only in the first row of $\tensor{M}$, hence its determinant is linear, as well. Following the findings of \textcite{yue1994plane, pence1991buckling} (see e.g. Fig. (4) in both papers), and employed later by \textcite{sunFoldingWrinklesThin2011}, among the possible values satisfying Eqn. \eqref{eq:system-condition}, only the ones corresponding to the smallest wavenumber are of interest. This leads to writing the following \textit{optimality conditions}, namely the ones governing both the minimum strain at  which the onset of wrinkling occurs and the corresponding wavenumber:
\begin{equation}
	\begin{cases}
		\varepsilon_{cr} = \hat{\varepsilon}(k_{h,cr}, \lambda_1^{cr}, \rho_{FM},\rho_{ML}, \kappa, \theta,k_2) \\
		\dfrac{\partial \varepsilon_{cr}}{\partial k_{h,cr}} = 0.
		\label{eqn:system-solution}
	\end{cases}
\end{equation}
Indeed, in full analogy with \textcite{sunFoldingWrinklesThin2011}, the conditions above can be shown to deliver the critical stretch and the corresponding non-dimensional wavenumber at which wrinkling occurs.

\noindent Due to the complexity of the OGH model, it is worthy of mention that the amount of calculations required to solve \eqref{eqn:system-solution} significantly increases relative to the case of neo-Hookean bilayers. Hence, a simpler model than OGH would be especially useful if it would deliver comparable results to its associated optimality conditions.

\noindent To this end, the OGH constitutive equation here is replaced by the Standard Reinforcing Model, SRM, mentioned above (for details see e.g. \parencite{Kurashige1981,TriantafyllidisAbeyaratne1983,Qiu1997RemarksOT}).
 The SRM energy density reads as follows:
\begin{equation}
	W_{\text{fiber,SRM}} = \dfrac{\gamma}{2}\sum_{m=1}^2E_{m}^2,
	\label{eq:energy-std-fiber-model}
\end{equation}
where $\gamma$ has the dimension of an elastic modulus. Of course, \eqref{eq:energy-std-fiber-model} must be added to the neo-Hookean term, accounting for the hyperelasticity of the matrix. 
It should be noted that, as the parameter $k_2$ approaches zero, the derivative of $W_{\text{fiber},OGH}$ with respect to the strain invariant $E_m$ coincides with the one of $W_{\text{fiber,SRM}}$ when $\gamma = k_1$ . 

\begin{figure}[tb]
	\centering
	\includegraphics{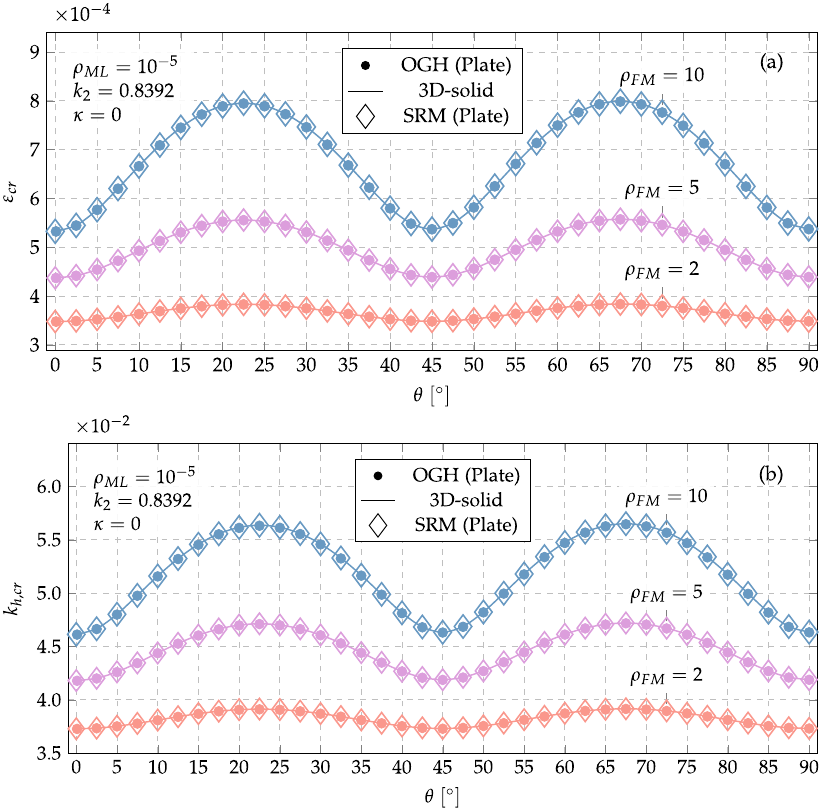}
	\caption{Comparison between the critical strain and the non-dimensional wavenumber between the plate model (dots) and the 3D-solid one (solid line) with respect to the angle $\theta$ and for three different values of $\rho_{FM}$. The considered ratio $\rho_{ML} = 10^{-5}$ is in the middle of the range of interest. The diamonds are the critical strains obtained from the Standard Reinforcing Model \eqref{eq:energy-std-fiber-model}.}
	\label{fig:comparison-solid-plate}
\end{figure}

\noindent Solutions of \eqref{eqn:system-solution} obtained by utilizing the SRM strain density energy are shown in Figure~\ref{fig:comparison-solid-plate}. From there, it is manifest that the critical strains are practically the same by using either constitutive equation, thereby suggesting that the assumption of a simpler constitutive law, such as the SRM, leads indeed to comparable results. This outcome is related to the independence of SRM from the parameter $k_2$. As a confirmation of this circumstance, \textcite{nguyenWrinklingInstabilitiesBiologically2020} illustrated that the OGH law does not depend on $k_2$ in the small strain regime: in the sequel, (see Figures~\ref{fig:comparison-solid-plate}$\div$~\ref{fig:asymptotic-kh-rhoML-e-4}) the magnitude of the arising strains are shown to be small enough.

\noindent A comparison between the outcomes of (i) the dimensionally reduced model coupled with SRM for the substrate and (ii) the solid one developed by \textcite{nguyenWrinklingInstabilitiesBiologically2020}, is shown in Figure~\ref{fig:comparison-solid-plate}. There, the critical strains and non-dimensional wavenumbers are displayed as functions of the angle formed by the two families of fibers (displayed in Figure 2) and for different stiffness ratio $\rho_{FM}$.  
This has been done by setting, as in \textcite{nguyenWrinklingInstabilitiesBiologically2020}, $k_2 = \num{0.8393}$ and $\kappa=0$. Moreover, $\nu_{L}=\num{0.5}$  has been imposed for the incompressibility of the layer and a stiffness ratio $\rho_{ML}=10^{-4}$ between the matrix and the layer has been assumed. The diagrams show that the results are essentially the same as the original model for exceptionally small stiffness ratios. Furthermore, a symmetric behavior for both the critical strain and the corresponding wavenumber about $\theta=\ang{45}$ is detected in the assumed range $10^{-6}\div 10^{-4}$ for $\rho_{ML}$, for the considered values of $\rho_{FM}$ and $\kappa$, namely the modified stiffness ratio between fibers and matrix (accounting for the volume concentration of the former) and the spatial dispersion of the fibers themselves. 

\section{A new asymptotic law for fiber-reinforced bilayers}
	\label{sec:an-asymptotic-law}
	The solution given by the system \eqref{eqn:system-solution} is not generally available in a simple form and, for a given set of the model parameters, it is therefore necessary to solve it numerically. However, it is still possible to simplify its expression under the following assumptions:
\begin{enumerate}[label=(\alph*)]
	\item\label{item:a1} the upper layer is considerably stiffer than the substrate, namely $\rho_{ML}=6\mu_M / E_L  \le10^{-4}$;
	\item the fibers ratio $\rho_{FM} = k_1 / \mu_M$ assumes values between 0 and 10. This hypothesis, although it may appear limiting as it would bring back to what has been already obtained through the neo-Hookean model, produces reliable results when (a) is fulfilled;
	\item the critical strain is very small ($\varepsilon_{cr}\le10^{-3}$, see e.g. \textcite{sunFoldingWrinklesThin2011} in the absence of fibers) and therefore, as $\varepsilon_{cr} = 1 - \lambda_1^{cr}$, $\lambda_1^{cr} \approx 1$. Because no prestretch is considered, the eigensolutions given by \eqref{eqn:eigensolutions} are linearized around $\lambda_1^{cr}=1$. 
    In this way, the resulting quantities, namely the critical strain and non-dimensional wavenumber, will have no dependence on the stretch;
	\item the critical strain and non-dimensional wavenumber are approximately constant when $k_2$ changes (at least for $\rho_{ML} = 10^{-4}$), as remarked in the previous section. It is thereby possible to replace the contribution of fibers given by the OGH constitutive law \eqref{eq:OGH-strain-energy} with the SRM \eqref{eq:energy-std-fiber-model}, removing the variable $k_2$ and the whole exponential part associated to that;
	\item the Poisson ratio of the layer is $\nu_L=\nicefrac{1}{2}$.
\end{enumerate}

\noindent By analogy with \textcite{allenAnalysisDesignStructural1969} and \textcite{sunFoldingWrinklesThin2011}, as pointed out in the previous section, explicit solutions to the optimality problem are sought. In other words, reliable asymptotic expansions for (i) the minimum of the critical strain yielding the onset of wrinkling and (ii) its corresponding wavenumber are the targets of this section.\\
\noindent In order to do so, one can start by taking advantage of the assumption \ref{item:a1}, i.e. $\rho_{ML} \approx 0$. Henceforth, a Taylor expansion of $(\partial\,\varepsilon_{cr}) / (\partial \, k_{h,cr})$ (the second equation appearing in \eqref{eqn:system-solution}) with respect to $\rho_{ML}$ around zero can be considered. 
Upon equating the obtained expression to zero, it is not difficult to check that a closed-form solution of an algebraic third-order equation in $k_{h,cr}$ can be obtained. The only possible physically admissible root of such an equation reads as follows:
\begin{gather}
	k_{h,cr} = \hat{k}_{h,cr}(\rho_{ML}, \rho_{FM}, \theta, \kappa) = \Big(\hat{k}_{h,cr}(\rho_{ML}, \rho_{FM}, \theta, \kappa)^3\Big)^{1/3} \approx \Big(g_1(\rho_{FM}, \theta, \kappa) \rho_{ML}\Big)^{1/3}, \label{eq:dimensionlesskexpansion}\\
    \text{where }g_1 = \bigg(\dfrac{\partial k_{h,cr}^3}{\partial \rho_{ML}}\bigg) \bigg|_{\rho_{ML}=0}. \notag
	\label{eq:asymptotic-law-kh}
\end{gather}
It is worth noting that the zero-order term of the expansion \eqref{eq:dimensionlesskexpansion} vanishes. From the physical viewpoint, this can be interpreted as the substrate becoming extremely soft relative to the thin top layer, when the matrix-to-film stiffness ratio $\rho_{ML} \to 0$. Hence, the buckling strain of such layer (for a finite depth of one unit length) of thickness $h$ turns out to be $\epsilon_{cr} \sim (h/L_0)^2$, as the wavelength tends to the physical length of the film. The thinness of the latter implies $h/L_0 \to 0$, hence both the critical strain and the corresponding dimensionless wave number, $k_{h,cr} \big|_{\rho_{ML} \to 0} \sim h/ L_0$, tend to zero.
It is worth noting that $g_1$ depends upon $\rho_{FM} = k_1 / \mu_M$, relating the stiffness of the fibers, weighted against their volume concentration, and the shear modulus of the substrate. For low densities of fiber reinforcements $\rho_{FM}$ tend to zero. Therefore, $g_1$ can be replaced by a suitable expansion obtained as follows:
\begin{equation}
	\begin{gathered}
		g_1(\rho_{FM}, \theta, \kappa) = \sqrt{g_1^2(\rho_{FM}, \theta, \kappa)} \approx \bigg(h_{10}(\theta, \kappa) + \rho_{FM} h_{11}(\theta, \kappa) + \dfrac{\rho_{FM}^2}{2} h_{12}(\theta, \kappa)\bigg)^{1/2},\\
        \text{where } h_{10} = g_1^2 \bigg|_{\rho_{FM}=0}, \quad
        h_{11} = \bigg(\dfrac{\partial g_1^2}{\partial \rho_{FM}} \bigg) \bigg|_{\rho_{FM}=0}, \quad
        h_{12} = \bigg(\dfrac{\partial^2 g_1^2}{\partial \rho_{FM}^2}\bigg) \bigg|_{\rho_{FM}=0} 
	\end{gathered}
\end{equation}
where $h_{1j}(\theta, \kappa),\; j=0,1,2$ are the terms of the expansions. Note that such expressions are valid only if $g_i$ are positive functions for every value of $\rho_{FM}$ and for $0\le\theta\le\pi/2$. This is reasonable since the wavenumber is a positive quantity. Finally, carrying out the computations of the previous expressions, one has
\begin{equation}
	\begin{gathered}
		h_{10} = 9,\quad
		h_{11} = 9(1-3\kappa)^2,\quad
		h_{12} = \dfrac{9\sin^2(4\theta)}{2}(1-3\kappa)^4.
	\end{gathered}
\end{equation}
Henceforth, the resulting critical (non-dimensional) wavenumber takes the following form:
\begin{equation}
	k_{h,cr} \approx \sqrt[3]{3\rho_{ML}} \sqrt[6]{1 + \rho_{FM}(1-3\kappa)^2 + \dfrac{\rho_{FM}^2 \sin^2(4\theta)}{4}(1-3\kappa)^4}
	\label{eqn:asymptotic-law-kh}
\end{equation}
and, as was previously pointed out, this value is unique.

\noindent A corresponding asymptotic expansion for the wrinkling strain can also be obtained. Indeed, by substituting \eqref{eqn:asymptotic-law-kh} in the first equation of \eqref{eqn:system-solution}, and by computing the Taylor expansion of the resulting expressions up to second order, the following form for $\varepsilon_{cr}$ is achieved:
\begin{equation}
\label{eqn:eps-crit-t0-t1-t2}
    \begin{gathered}
	\varepsilon_{cr} = \Big(\hat{\varepsilon}_{cr}(\rho_{FM},\rho_{ML},\theta, \kappa)^3\Big)^{1/3} \approx \left(t_2(\rho_{FM}, \theta, \kappa) \dfrac{\rho_{ML}^2}{2}\right) ^{1/3},\\
    \text{where }t_2 = \bigg(\dfrac{\partial^2 \varepsilon_{cr}^3}{\partial \rho_{ML}^2}\bigg) \bigg|_{\rho_{ML}=0}.
    \end{gathered}
\end{equation}
Similarly to the previous case, the zero-order term in the Taylor expansion for the argument in \eqref{eqn:eps-crit-t0-t1-t2} is zero and, furthermore, in this specific case, even the first-order one identically vanishes.  In particular, the zero-order term corresponds to $\varepsilon_{cr}\big|_{\rho_{ML}=0}$, which is again consistent with having a compressed free-standing (because $\rho_{ML}=0$ would essentially mean to have a substrate with zero stiffness relative to the top layer) infinitely thin film with a finite length.\\
Furthermore, in order to achieve an irreducible representation for the critical strain, a Taylor expansion of $t_2$ can be provided. By expanding that with respect to $\rho_{FM}$ around $0$, the following expression follows:
\begin{equation}
	\begin{gathered}
		t_2(\rho_{FM}, \theta, \kappa) \approx q_{20}(\theta, \kappa) + \rho_{FM} q_{21}(\theta, \kappa) + \dfrac{\rho_{FM}^2}{2} q_{22}(\theta, \kappa),\\
        \text{where } q_{20} = t_2^2 \bigg|_{\rho_{FM}=0}, \quad
        q_{21} = \bigg(\dfrac{\partial t_2^2}{\partial \rho_{FM}} \bigg) \bigg|_{\rho_{FM}=0}, \quad
        q_{22} = \bigg(\dfrac{\partial^2 t_2^2}{\partial \rho_{FM}^2}\bigg) \bigg|_{\rho_{FM}=0}. 
	\end{gathered}\label{eq:p2}
\end{equation}
Furthermore, by carrying out the computations for $q_{2i},\;i=0,1,2$, their values read as follows:
\begin{equation}
q_{20} = \dfrac{9}{32}, \quad 
q_{21} = \dfrac{9}{32}(1-3\kappa)^2, \quad
q_{22} = \dfrac{9\sin^2(4\theta)}{64}(1-3\kappa)^4.
\label{eq:q-terms}
\end{equation}
Upon substituting \eqref{eq:q-terms} into \eqref{eq:p2}, and the obtained result into \eqref{eqn:eps-crit-t0-t1-t2}, the following asymptotic expression for the critical strain is delivered:
\begin{equation}
	\varepsilon_{cr} \approx \dfrac{\sqrt[3]{\left(3 \rho_{ML}\right)^2}}{4} \sqrt[3]{1 + \rho_{FM}(1-3\kappa)^2 + \dfrac{\sin^2(4\theta)}{4}(1-3\kappa)^4}.
	\label{eqn:asymptotic-law-strain}
\end{equation}

\begin{figure}
    \centering
    \includegraphics[width=0.65\linewidth]{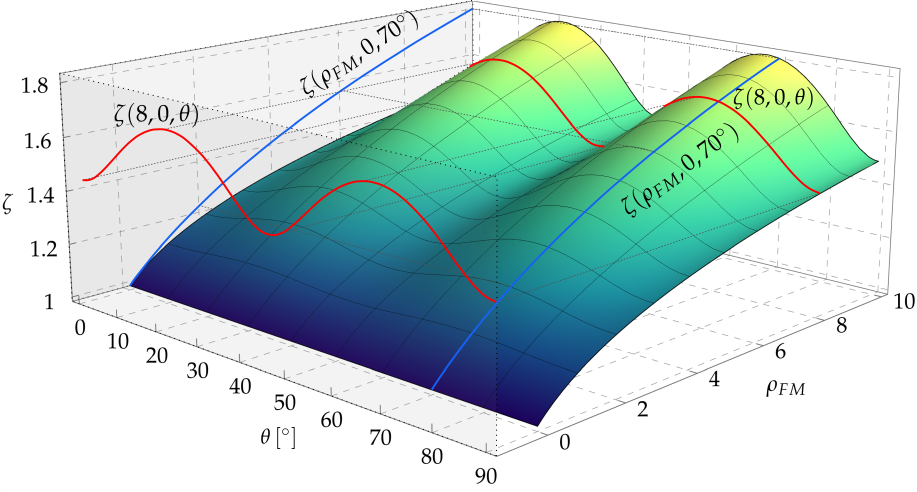}
    \caption{The amplitude function $\zeta(\rho_{FM}, \kappa, \theta)$ varying $\rho_{FM}$ and $\theta$, assuming $\kappa=0$ (perfect alignment). In particular, two projections are shown: the first one, within the plane $\rho_{FM}-\zeta$ shows the amplification for a fixed angle, namely $\theta=\ang{70}$. The second one, in the plane $\theta-\zeta$ represents the sinusoidal amplification by setting $\rho_{FM} = 8$. Note that when the dispersion factor $\kappa$ approaches zero the function is an horizontal plane with $\zeta=1$.}
    \label{fig:zeta-3D}
\end{figure}

\begin{figure}
	\centering
	\includegraphics{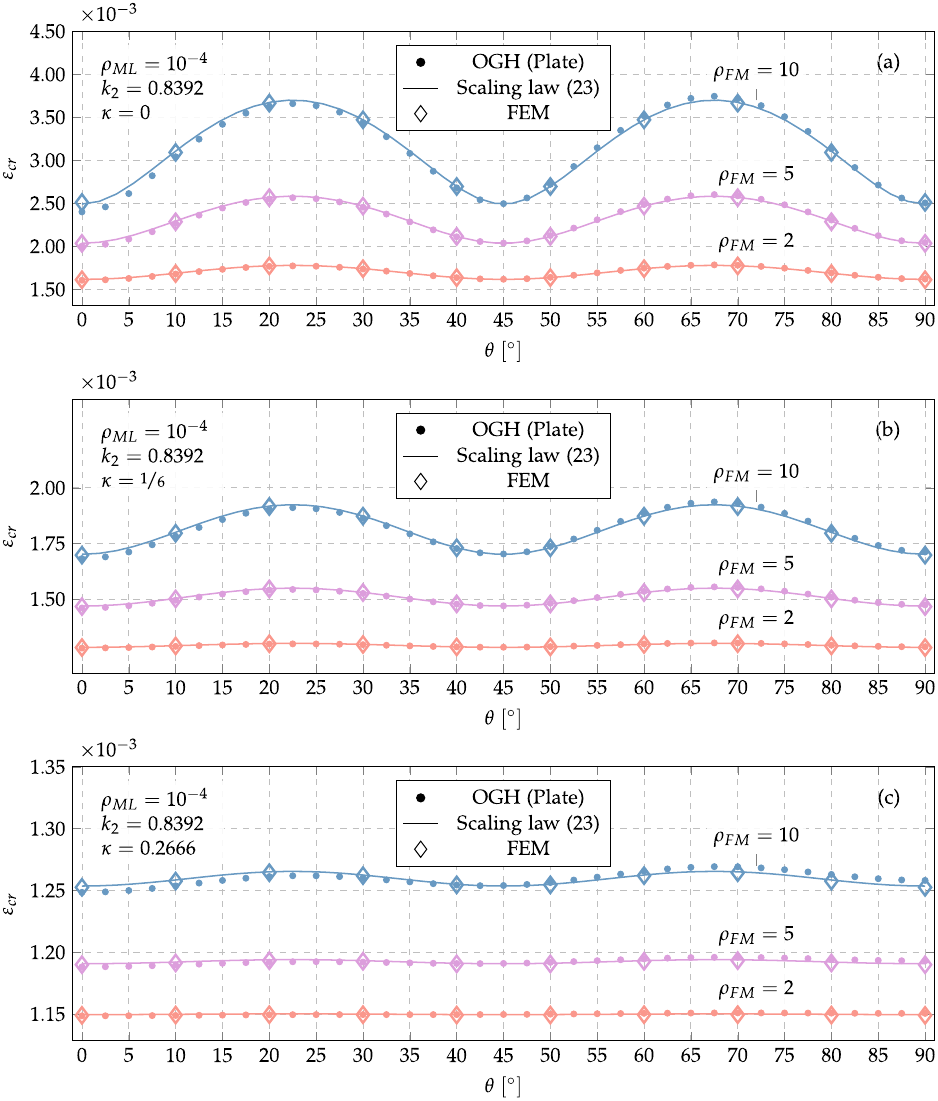}
	\caption{Comparison of the critical strain between FEM, the plate model (dots) and the asymptotic expansion \eqref{eqn:asymptotic-law-strain} (solid line) with respect to the angle $\theta$, $\rho_{ML}=10^{-4}$ and $\rho_{FM}=2,5,10$.}
	\label{fig:asymptotic-strain-rhoML-e-4}
\end{figure}

\noindent For the sake of consistency, \eqref{eqn:asymptotic-law-kh} and \eqref{eqn:asymptotic-law-strain} are explored for the simpler case of neo-Hookean bilayers.
As expected, those expressions reduce to what already found in \textcite{sunFoldingWrinklesThin2011} and \textcite{caoWrinklingPhenomenaNeoHookean2012} by assuming $\nu_{L} = 1/2$, i.e.
\begin{gather}
	k_{h,nh} = \sqrt[3]{3 \rho_{ML}}
	\label{eqn:asymptotic-law-kh-nh}\\
	\varepsilon_{cr, nh} = \dfrac{\sqrt[3]{\left(3 \rho_{ML}\right)^2}}{4}.
	\label{eqn:asymptotic-law-strain-nh}
\end{gather}
It is evident by inspections of \eqref{eqn:asymptotic-law-kh} and \eqref{eqn:asymptotic-law-strain} that the presence of the fibers turns out to significantly influence both the wrinkling strain and the corresponding wavenumber. Indeed, the asymptotic expressions above involve the following \textit{modulating factor}:
\begin{equation}
    \zeta(\rho_{FM}, \theta, \kappa) = \sqrt[6]{1+ \rho_{FM}(1-3\kappa)^2 + \dfrac{\rho_{FM}^2\sin^2(4\theta)}{4}(1-3\kappa)^4}.
    \label{aplifying_factor}
\end{equation}
Henceforth, the quantities mentioned above can be written as
\begin{gather}
    \label{eqn:asymptotic-law-kh-final}
	k_{h,cr} = k_{h,nh} \;\zeta(\rho_{FM}, \kappa, \theta)
	\\
	\varepsilon_{cr} = \varepsilon_{cr,nh} \;  \zeta^2(\rho_{FM}, \kappa,  \theta)
	\label{eqn:asymptotic-law-strain-final}
\end{gather}
where $\zeta(\rho_{FM}, \kappa, \theta)$ (see Figure~\ref{fig:zeta-3D}) is defined by the expressions above and its square has the meaning of amplitude factors for the neo-Hookean values of $k_h$ and $\varepsilon_{cr}$, respectively. It is just as simple to verify that the modulating factor is really an amplification function. This reduces to 1 in both cases in the absence of fibers, as shown in Figure~\ref{fig:zeta-3D}, and when $\kappa$ approaches to $1/3$. This demonstrates that \emph{in the case of a total dispersion of fibers within the matrix these give no contribution to the critical dimensionless wavenumber and strain}.
\noindent Using expressions \eqref{eqn:asymptotic-law-kh-final} and \eqref{eqn:asymptotic-law-strain-final}, and the results achieved in \eqref{eqn:system-solution}, the plots displayed in Figures~\ref{fig:asymptotic-strain-rhoML-e-4}~$\div$~\ref{fig:logarithmic-plots} for different angles, stiffness ratio $\rho_{ML}$, $\rho_{FM}$ and dispersion factors are obtained.
It is worth noting that Eqn.s \eqref{eqn:asymptotic-law-kh-final} and \eqref{eqn:asymptotic-law-strain-final} work particularly well for small $\rho_{ML}$ ratios: this can be achieved even for stiff fibers, provided that their concentration per unit volume is adequately low. %
Another aspect is related to the behavior of the curves: for small $\rho_{ML}$, the trends of both critical strain and wavenumber are symmetrical with respect to $\pi/4$.\\
The same plots display the results coming from the utilized FEM. The numerical simulations have been performed by means of the commercial software ABAQUS/Standard (Lic. n. LKO2211177).\\
The substrate is modeled as a three-dimensional hyperelastic body under plane strain conditions. In particular, within the plane of interest, a rectangle of length equal to ten times the wavelength $\lambda=2 \pi h/ k_h$ (provided by the asymptotic expansion \eqref{eqn:asymptotic-law-kh-final}) is considered.  For the sake of computation, the depth of the substrate is taken as the maximum between one hundred times the thickness of the top layer and two times the wavelength. This choice essentially provides a semi-infinite substrate compared to the thin layer, and it relies on a theoretical justification thanks to \textcite{yue1994plane, pence1991buckling} (see the barreling curves reported in such papers in Fig.s 4, same numbering in each paper).\\
Based on the geometry described above, it is clear that the smaller the ratio $\rho_{ML}$ is, the bigger the size of the substrate will be. Upon utilizing two-dimensional plane strain solid elements, the thickness of the layer governs the characteristic size of the mesh (that is to avoid distorted mesh that could lead to numerical ill-posedness of the governing operator). As a result, the simulations with small $\rho_{ML}$ will be penalized from a computational point of view, due to an extremely high number of nodes.\\
Therefore, the following choices have been adopted for the discretization and computational analysis of the system. As pointed out in \parencite{sunFoldingWrinklesThin2011}, instead of modeling the upper layer on the basis of a two-dimensional geometry, the thinness of the top layer relative to its length suggests the use of modified B22 beam elements, by means of the built-in stringer option in ABAQUS. The elements just mentioned above are indeed properly modified to represent the plate behavior of the layer undergoing plane strain conditions. The latter evidently constrains the lateral contraction/expansion of each stripe of the top film during wrinkling. Henceforth, the representative cross-section of the thin layer can be taken with unit height, whereas its base must be set equal to $(1-\nu_L^2)^{-1}$. Regarding the substrate, eight-node, hybrid, plane strain elements CPE8H are adopted. In order to properly display the wrinkling mode, the mesh size is calibrated at about $\lambda_{cr}/40$.\\
The outcomes of the numerical simulations performed on the model are in excellent agreement with the analytic ones for $\rho_{ML}=10^{-6},10^{-5}$. Indeed, the error between the analytic and the computational results are of the order of $10^{-3}$. Upon exploring cases for which
 $\rho_{ML}=10^{-4}$, such an error rises to $10^{-2}$, thereby suggesting that lowering this discrepancy can be done by modeling the film itself through the CPE8H elements mentioned above.

\begin{figure}
	\centering
	\includegraphics{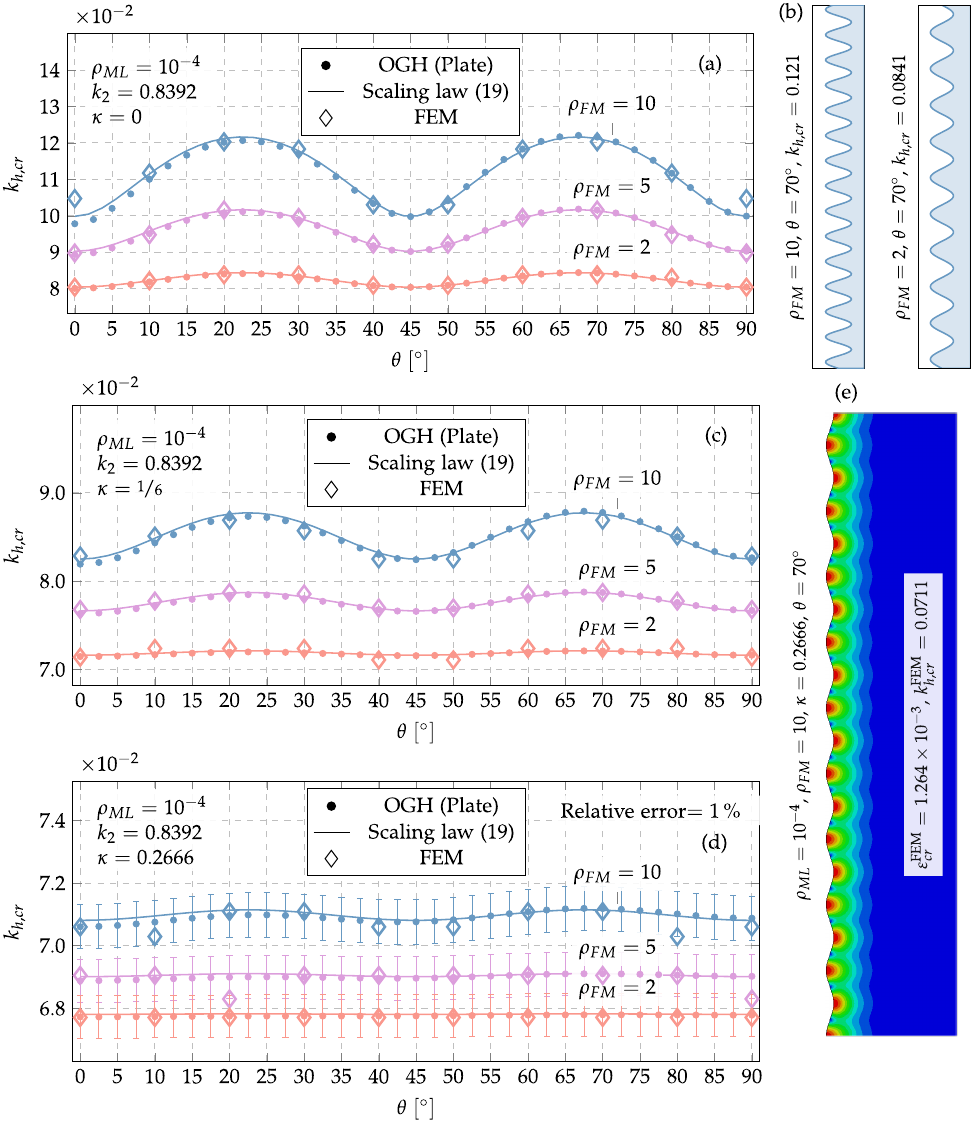}
	\caption{Comparison of the critical dimensionless wavenumber between FEM, the plate model (dots) and the asymptotic expansion \eqref{eqn:asymptotic-law-kh} (solid line) with respect to the angle $\theta$, $\rho_{ML}=10^{-4}$ and $\rho_{FM}=2,5,10$ ((a), (c), (d)). In (b) the wrinkling mode, with normalized amplitude, of a representative set of values is shown. Finally, in (e) the magnitude of the normalized displacements resulting from the FE analysis is plotted.}
	\label{fig:asymptotic-kh-rhoML-e-4}
\end{figure}

\begin{figure}
    \centering
	\includegraphics{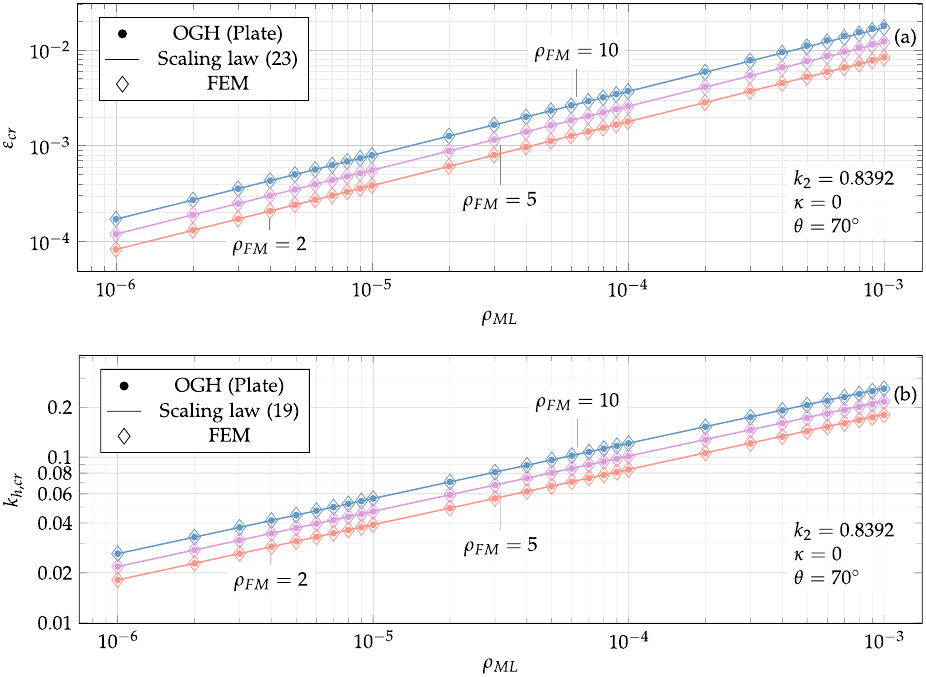}
    \caption{Comparison of the critical strain and wavenumber for $\kappa=0$, $\theta=\ang{70}$ and $k_2=\num{0.8393}$, as a function of the stiffness mismatch between the matrix and the film $\rho_{ML}$. Note that the plots are on a bi-logarithmic scale.}
    \label{fig:logarithmic-plots}
\end{figure}

\noindent Concerning the material properties, while the layer is described by a classical linear elastic law with Poisson ratio equal to $1/2$ (because of incompressibility), for the substrate a custom UMAT routine to properly simulate the OGH constitutive relation defined by \eqref{eq:OGH-strain-energy} has been written. This is because the built-in ABAQUS routine neglects the contribution of the compressed fibers, deactivating them once they buckle.\\
Finally, an extended buckling analysis is performed by making use of an extended number of simulations (actually over \num{170}), due to the need to cover the whole range of parameters.
%
From Figures~\ref{fig:asymptotic-strain-rhoML-e-4}~$\div$~\ref{fig:logarithmic-plots} it is clear how the outcomes of FEM 
display a full agreement with theoretical ones. In particular, Figures~\ref{fig:asymptotic-strain-rhoML-e-4} and \ref{fig:asymptotic-kh-rhoML-e-4} show the critical strain and wavenumber with respect to the angle for $\rho_{ML}=10^{-4}$ and different values of mismatch fibers/matrix $\rho_{FM}$. It is worth noting that, the closer to $1/3$ the dispersion is, the flatter the curves are, approaching to the neo-Hookean case when $\kappa=1/3$ (as well as when $\rho_{FM}=0$). Finally, in Figure~\ref{fig:logarithmic-plots} the critical quantities are shown by setting a constant angle $\theta=\ang{70}$ and varying the ratio $\rho_{ML}$, assuming a perfect alignment of the fibers.
Furthermore, it is noteworthy that the representation of the critical strain and wavenumber with respect to the stiffness mismatch $\rho_{ML}$ is susceptible to an interesting property. In fact, by using a bi-logarithmic scale as in Figure~\ref{fig:logarithmic-plots}, it becomes apparent that these quantities arrange along straight lines with slope $2/3$ and $1/3$, respectively.

\begin{figure}
	\centering
	\includegraphics[width=0.95\linewidth]{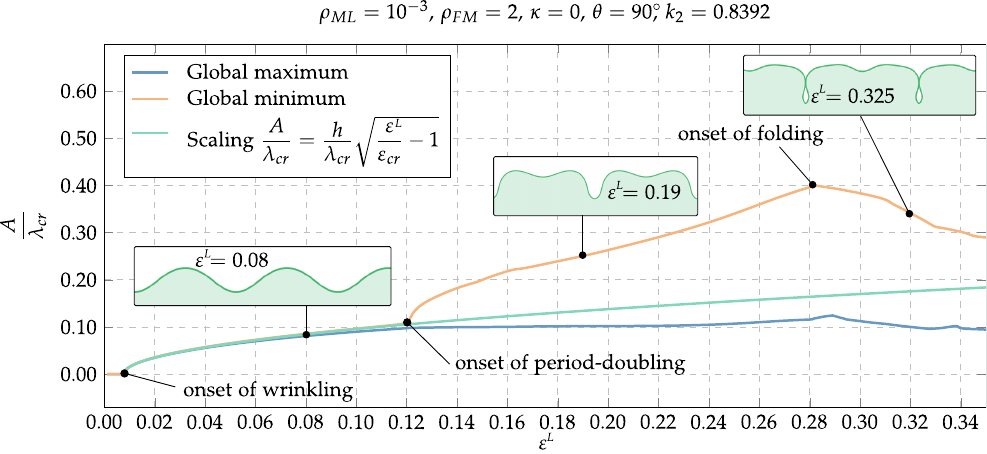}
	\caption{Dimensionless amplitude $A/\lambda_{cr}$ for the case $\rho_{ML}=10^{-3}$, $\rho_{FM}=2$, $k_2=\num{0.8393}$, $\kappa=0$ and $\theta=\ang{90}$. As the contractile strain increases different patterns emerge on the surface. Initially, the film is flat but, after reaching the critical strain $\varepsilon_{cr}$ wrinkling occurs. In such a region the amplitude turns out to scale like $h \sqrt{\varepsilon/\varepsilon_{cr} - 1}$. In the post-wrinkling regime period-doubling and creases can be observed.}
	\label{fig:amplitude}
\end{figure}

\noindent For the sake of completeness, in Figure \ref{fig:amplitude} the post-buckling phase of a bilayer is shown. For this case $\rho_{ML}=10^{-3}$, $\rho_{FM}=2$, $k_2=\num{0.8393}$, $\kappa=0$ and $\theta=\ang{90}$ have been assumed. The amplitude has been evaluated as the absolute value of the deviation from the mean height of the surface of two representative points, namely the global maximum and minimum.\\
Figure \ref{fig:amplitude} shows that, depending on the contractile strain, the dimensionless amplitude defined by $A/\lambda_{cr} = A k_{cr}/(2 \pi)$ changes, and a re-organization of the surface emerges. Indeed, while the upper surface is initially flat, after the onset of the wrinkling the amplitude increases, with the current wavelength being $k_{cr}/\lambda_{1}^{cr}$. The field \eqref{eqn:definition-p} reproduces the kinematics that the bilayer has until the onset of the period-doubling. Note that in such a region, which extends up to ten times the wrinkling strain, the amplitude is governed by the well-known relation $A = h\sqrt{\varepsilon / \varepsilon_{cr} - 1}$. This has been obtained in the absence of fibers, as reported by \textcite{huangNonlinearAnalysesWrinkles2005a, chenHerringboneBucklingPatterns2004, maneRatedependentWrinklingSubsequent2022} among others. Moreover, beyond a certain strain, two consecutive crests begin to join, and the valley between them flattens out, causing period-doubling. Finally, by further increasing the compression the waves move closer, until they make contact with one another (folding).

\section{Approximation based on linearized orthotropic properties}
 	\label{sec:orthotropic}
 	In order to explain the trend of the critical strain an approximation has been constructed in \textcite{nguyenWrinklingInstabilitiesBiologically2020}
by treating the substrate as an orthotropic material. This leads to the derivation of stiffness moduli and Poisson’s ratios through an appropriate linearization.
In this section it is proved that the scaling laws \eqref{eqn:asymptotic-law-kh-final} and \eqref{eqn:asymptotic-law-strain-final} can actually be rewritten by taking into account the approximation above, thereby expressing them in terms of stiffness parameters of the linearized orthotropic substrate.

\noindent In a fiber-reinforced material, the principal directions $P_1-P_2$ are identified by the orientation of fibers and by their normal, which may not coincide with the "natural" reference system $X_1-X_2$ used to solve the equilibrium problem. Using the expressions found by \textcite{nguyenWrinklingInstabilitiesBiologically2020}, the stiffness moduli with respect to the natural directions result as follows
\begin{equation}
	\begin{gathered}
		E_{X_1} = \mu_M\dfrac{6 + 8 \rho_{FM} (1-3\kappa)^2(1-3\cos^2(\theta)\sin^2(\theta))}{1 + \rho_{FM}(1-3\kappa)^2\sin^4(\theta)}\\
		E_{X_2} = \mu_M\dfrac{6 + 8 \rho_{FM} (1-3\kappa)^2(1-3\cos^2(\theta)\sin^2(\theta))}{1 + \rho_{FM}(1-3\kappa)^2\cos^4(\theta)}.\\
	\end{gathered}
\end{equation}
Henceforth, their value in the principal system is obtained by placing $\theta=0$, by assuming the axis $P_1$ and $X_1$ aligned, so that
\begin{equation}
	\begin{gathered}
		E_{P_1} = E_{X_1}(\theta=0) = \mu_M\big(6 + 8\rho_{FM}(1-3\kappa)^2\big),\\
		E_{P_2} = E_{X_2}(\theta=0) =\dfrac{\mu_M\big(6 + 8\rho_{FM}(1-3\kappa)^2\big)}{1 + \rho_{FM}(1-3\kappa)^2}. 
	\end{gathered}
\end{equation}
Noting that their ratio is
\begin{equation}
	\dfrac{E_{P_1}}{E_{P_2}} = 1 + \rho_{FM}(1-3\kappa)^2
\end{equation}
it is clear it can be substituted into the amplitude function \eqref{eqn:asymptotic-law-kh-final}, obtaining
\begin{equation}
	\zeta(\rho_{FM}, \kappa, \theta) = \zeta_{\text{ortho}}\big(E_{P_1}/E_{P_2}, \theta\big) = \sqrt[6]{\dfrac{E_{P_1}}{E_{P_2}} + \Bigg(\dfrac{E_{P_1}}{E_{P_2}} - 1\Bigg)^2\dfrac{\sin^2(4\theta)}{4}}.
\end{equation}
In this way the scaling laws \eqref{eqn:asymptotic-law-kh-final} and \eqref{eqn:asymptotic-law-strain-final} can be rewritten as follows
\begin{equation}
	k_{h, cr}^\text{ortho} = k_{h,nh}\;\zeta_{\text{ortho}}\big(E_{P_1}/E_{P_2}, \theta\big),
	\quad
	\varepsilon_{cr}^{\text{ortho}} = \varepsilon_{h,nh}\;\zeta_{\text{ortho}}^2\big(E_{P_1}/E_{P_2}, \theta\big)
\end{equation}
from which one can see how the response depends on the ratio of the orthotropic stiffness moduli in the principal system and on the angle that fibers have with respect to the natural one.

\begin{figure}
    \centering
	\includegraphics{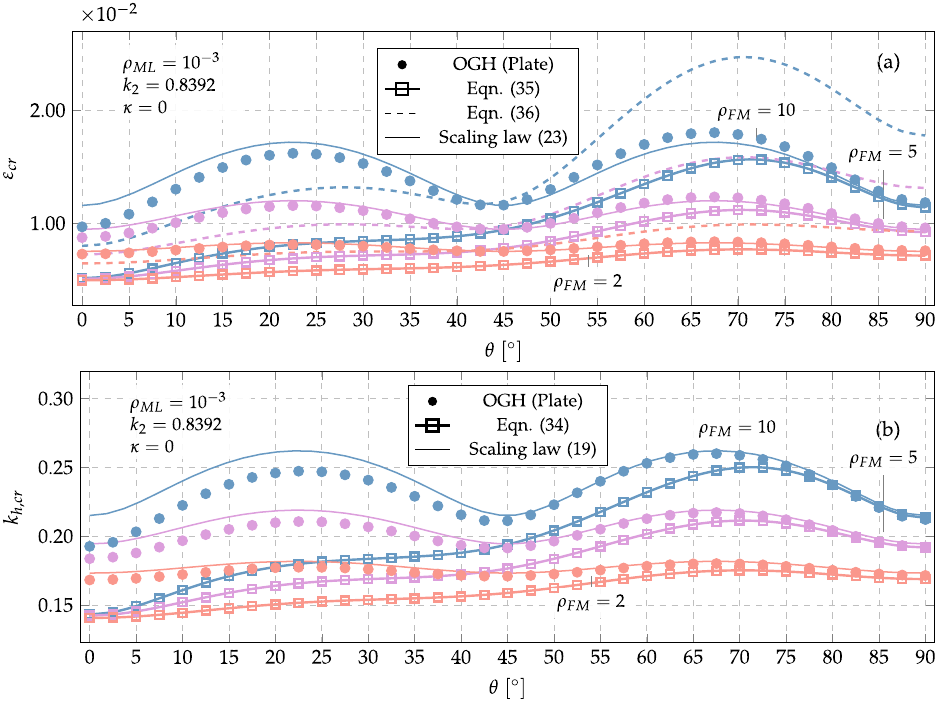}
    \caption{Comparison of the critical strain (a) and dimensionless wavenumber (b) assuming $\rho_{ML}=10^{-3}$, $\kappa=0$ and $k_2=\num{0.8393}$. Although the curves are slightly asymmetric for $\rho_{FM}=10$, it emerges that scaling laws \eqref{eqn:asymptotic-law-kh-final} and \eqref{eqn:asymptotic-law-strain-final} approximate the exact results better than formulations proposed in \textcite{vonachEffectsInplaneCore2000} and in \textcite{nguyenWrinklingInstabilitiesBiologically2020}.}
    \label{fig:comparison-orthotropic}
\end{figure}

By assuming a bilayer with a geometry similar to what is considered in this paper, though with a linear elastic orthotropic substrate instead of a fiber-reinforced one, \textcite{vonachEffectsInplaneCore2000} obtained scaling laws for such systems.  
In Eqns. (22) and (23), such authors provided explicit formulas for the semi-wavelength and the critical longitudinal load at the onset of wrinkling of an isotropic thin plate resting on an elastic foundation. By adapting these results to write down the actual wavelength and the critical strain it follows that
\begin{gather}
    k_{h,cr} = h \sqrt[3]{\dfrac{k_s}{2 K_L}} \label{eqn:V-R-k}\\
    \varepsilon_{cr} =  \dfrac{h^2}{12} \left(\dfrac{1}{\sqrt[3]{4}} + \sqrt[3]{2}\right)\sqrt[3]{\dfrac{k_s^2}{K_L^2}}, \label{eqn:V-R-e}
\end{gather}
where $K_L = E_L \, h^3/\left(12(1-\nu_L^2)\right)$ is the bending stiffness of the top layer, while $k_s$ is the substrate stiffness. It is noteworthy that these relations are particularly similar to the ones valid for isotropic bilayers. Depending on the specific problem under consideration, $k_s$ can assume different forms. For instance, the outcomes of Eqn.s \eqref{eqn:V-R-k} and \eqref{eqn:V-R-e} displayed in Figure \ref{fig:comparison-orthotropic}, arise by choosing the substrate's stiffness $k_s$ as in Eqn. (21) of \textcite{vonachEffectsInplaneCore2000}. It must be mentioned that for the evaluation of such a quantity the linearized orthotropic moduli obtained in Appendix 2 of \textcite{nguyenWrinklingInstabilitiesBiologically2020} have been used. Furthermore, in  Eqn. (7) of that same paper, the authors constructed a scaling law for the critical strain that fits the results obtained for low substrate/layer mismatches with good agreement. Such an expression is neither based on an asymptotic expansion nor on a rigorous derivation, and it makes use of an ad-hoc elastic module, $E_{\text{eff}}^s$, such that the following relations hold:
\begin{equation}
    \varepsilon_{cr} = \num{0.85} \sqrt[3]{\dfrac{E_{\text{eff}}^s G_{xy}^s}{E_{L}^2/(1-\nu_L^2)^2}}, \quad \text{with } E_{\text{eff}}^s \approx \dfrac{\sqrt{E_x^s E_y^s}}{1-\nu_{xz}^s\nu_{zx}^s}.
\end{equation}
Finally, unlike Eqn. (22) and (23) by \textcite{vonachEffectsInplaneCore2000} and the Eqn. (7) by \textcite{nguyenWrinklingInstabilitiesBiologically2020}, it is worthy of mention that Figure \ref{fig:comparison-orthotropic} shows how scaling laws \eqref{eqn:asymptotic-law-kh-final} and \eqref{eqn:asymptotic-law-strain-final} derived in this paper well capture the quasi-symmetric trend of the critical wavenumber and critical strain at the onset of wrinkling.
Indeed, the case $\rho_{ML}=10^{-3}$ portrayed in  Figure \ref{fig:comparison-orthotropic} presents a slight asymmetry when $\rho_{FM}=10$ that is not present whenever $\rho_{ML} \le 10^{-4}$ and $0 \le\rho_{FM} \le 10$.\\
Therefore, in the range of parameters examined in this present work, the novel scaling laws \eqref{eqn:asymptotic-law-kh-final} and \eqref{eqn:asymptotic-law-strain-final} show minor deviations from the analytical model, and they are definitely much closer to the exact results compared to the other formulations available in the literature.

\section{Conclusions}
    \label{sec:conclusions}
	Compressed bilayers made of stiff thin films perfectly bonded to the top of fiber-reinforced deep soft substrates have been studied in this paper. To begin with, it has been shown how assuming a plate behavior for the thin top layer actually allows the reproduction of similar results to the ones obtained in \textcite{nguyenWrinklingInstabilitiesBiologically2020}, where a three-dimensional elastic behavior for the film itself was adopted. It has then been illustrated that for high stiffness mismatch ratios between the substrate and the adhering layer, both models essentially show the same results, although the dimensional-reduced model naturally entails significantly less computational cost. 
\noindent Upon varying the relative angle between the two families of fibers, the simulations performed for cases in which the substrate is much softer than the layer yield a symmetric and sinusoidal trend both for the critical strain and for the corresponding wavenumber. 

\noindent Although complex theoretical and numerical analyses can be performed case-by-case, a prompt evaluation of the main wrinkling features for the bilayers at hand is often needed. 
This is certainly the case when it comes to comparing experimental findings with handy estimates of the topographic features of the corrugation in such systems. Unfortunately, no tools yet exist owing to the rapid evaluations mentioned above.
To this end, appropriate \textit{scaling laws} would certainly fill such a gap.
In particular, no rigorously derived simple relations governing the critical strain and the wave number at the onset of wrinkling had been provided for highly mismatched hyperelastic fiber reinforced bilayers before this present work.
This drawback did include biological systems, even in the case of estimating the most basic features of the exhibited corrugated topography of organs such as arteries. Although in \textcite{nguyenWrinklingInstabilitiesBiologically2020} several interesting analyses were performed (actually by a team of researchers involving three of the coauthors of this present paper) for flat fiber reinforced bilayers, a systematic rationale for mathematically deriving scaling laws for the features highlighted above was not pursued. Indeed, in such a paper only an ad-hoc scaling for the strain at the onset of wrinkling was provided for specific cases.

\noindent The main result of this work relies upon the \textit{novel scaling laws} for the critical strain and its corresponding wavenumber characterizing the initial wrinkling of compressed fiber-reinforced bilayers. This has been achieved only thanks to the outcomes of the simplified model developed in this paper. Indeed, close-form solutions to the bifurcation problem governing the balance of forces at the interface between the film (treated as a thin plate) and the fiber-reinforced substrate (modeled as an SRM) enabled one  to analytically find explicit asymptotic expansions owing to the new scalings mentioned above. Remarkably, either the absence of fibers or their complete randomness reduce the new scaling laws to the well-known ones valid for neo-Hookean bilayers (see e.g. \parencite{sunFoldingWrinklesThin2011,caoWrinklingPhenomenaNeoHookean2012}). 

\noindent In all the other cases, a \textit{modulating factor} depending on the properties of the fibers turns out to govern the newly obtained laws (see Figure \ref{fig:zeta-3D}). 
As expected, for certain fiber orientations and for certain fiber-matrix stiffness ratios, this can amplify the (dimensionless) wavenumber up to a factor $1.80$ and the corresponding wrinkling strain of over $3.2$. This significant amplification would then be missing if the novel scaling laws were erroneously replaced either by existing estimates based on the film and the matrix properties alone or on the ad-hoc relations mentioned above. 
%
\noindent The analyses in Sect.s $2 \div 3$, among other results, provide detailed comparisons between the outcomes of the novel scaling laws against both the results obtained from the dimensionally reduced analytical approach, and from the FE analyses. These results show how truly satisfactory the outcomes of the newly obtained scaling laws are relative to the corresponding analytic and numerical results. Furthermore, Sect. 4, Figure \ref{fig:comparison-orthotropic} displays the discrepancies between the results coming from fully linear approaches (i.e. \textcite{vonachEffectsInplaneCore2000}) and the ad-hoc relation found in \textcite{nguyenWrinklingInstabilitiesBiologically2020} versus both the novel scalings and the analytic formulation for certain values of the parameters of the fiber-reinforced bilayers. Unlike the first two sets of results, Figure \ref{fig:comparison-orthotropic} displays how the last two methods, which have been shown to essentially agree with the outcomes of the FE analyses in the previous sections, reliably hold throughout the whole range of variability of the fiber's angle.

%
\noindent Finally, in Sect. 5 it has also been shown how material parameters obtained in \textcite{nguyenWrinklingInstabilitiesBiologically2020}, appendix 2, providing orthotropic linearized moduli for the substrate 
can be used to express the newly obtained scaling laws. 

\noindent Providing tools like appropriate and rigorously derived scaling laws is key for the analysis of geometrically complex situations involving compressed highly mismatched bilayers containing fibers, with whatever degree of dispersion relative to a main orientation they have.
Indeed, \textit{having the availability of scaling laws for the main features of wrinkling}, i.e. the critical strain at its onset and the associated wavenumber, \textit{becomes definitely useful when it comes to performing a firsthand comparison with experimental measurements even before running computational/analytical analyses}.\\   
The strategy undertaken in this paper to obtain such laws is currently under investigation for low-mismatch fiber-reinforced bilayers, that are even more amenable for soft biological tissues. Furthermore, new scaling laws in the presence of possible sources of inelasticity, such as growth, or even yielding and plasticity of the top layer in the presence of metallic films may be accounted for upon generalizing the methodology proposed in this paper to such situations. In other words, the systematic approach introduced in this paper to asymptotically expand complex representation formulas of the main wrinkling features paves the way for many other related problems, such as pre-stretch induced corrugation both in flat and curved fiber-reinforced bilayers, the latter being much closer to the geometry of the cross sections of wrinkled arteries.
 
\section*{Acknowledgements} 
A.M. and L.D. gratefully acknowledge the Italian Ministry of Universities and Research (MUR) in the framework of the project DICAM-EXC, Departments of Excellence 2023-2027 (grant DM 230/2022).\\
L.D., A.C. and M.F. gratefully acknowledge the partial support from the grants:   Italian Ministry of Universities and Research (MUR) through the grants PRIN-20177TTP3S and PON “Stream”-ARS01 01182.\\
L.D.and A.C. gratefully acknowledge the partial support from the grant PRIN-2022XLBLRX.\\
M.F. gratefully acknowledges the partial support from the grant PRIN-2022ATZCJN.\\
L.D. also gratefully acknowledges the partial support from the ERC through (1) FET Open “Boheme” grant no. 863179, (2) LIFE GREEN VULCAN LIFE19 ENV/IT/000213, (3) ERC-ADG-2021-101052956-BEYOND, (4) the Italian Group of Theoretical Mechanics GNFM-INDAM of the \emph{National Institute of High Mathematics}, and (5) the Italian Government through the 2023-2025 PNRR\_CN\_ICSC\_Spoke 7\_CUP E63C22000970007 grant, awarded to the University of Trento, Italy.\\
The authors wish to thank Dr. Thomas A. Witten, Homer J. Livingston Professor Emeritus within the Department of Physics and the James Franck Institute of the University of Chicago for the fruitful online conversations about the wrinkling of thin systems.


\printbibliography

\end{document}